\documentstyle[12pt]{article}
\begin{document}
\begin{flushright}
Preprint IHEP 96-2\\
hep-ph/9602347
\end{flushright}
\begin{center}
{\large\bf Hadronic production of $S$- and $P$-wave states of $\bar b
c$-quarkonium}\\
\vspace*{3mm}
A.V.~Berezhnoy, V.V.~Kiselev, A.K.~Likhoded\\
{\it Theory Division, Institute for High Energy Physics,\\
Protvino, Moscow Region, 142284, Russia.\\
E-mail: likhoded @mx.ihep.su\\
Fax: +7-095-230 23 37}
\end{center}
\begin{abstract}
In the leading $O(\alpha_s^4)$ order of the perturbative QCD, the hadronic 
production cross-sections of $S$- and $P$-wave states of $B_c$ meson are
calculated. The results for the $S$-wave levels are compared with the
values given in other papers as well as in the model of the $b$ quark
fragmenatation into $B_c$. In the given order, the cross-sections of 
the hadronic production of the $P$-wave states are calculated for the 
first time. Their contribution into the $B_c$ meson production is less than
10\%. There is a strong difference between the predictions of the
fragmentation model and the exact perturbative calculations. These
difeerences are discussed in details for the differential distributions
over variuos kinematical quantities.
\end{abstract}

\section*{Introduction}
During recent two years an essential progress has been reached in the
understanding of the production mechanisms for the heavy quarkonium,
composed of two heavy quarks with different flavours.

The $\bar b c$-quarkonium production process has the most simple 
consideration in $e^+e^-$-annihilation, where in the limit of
high energies ($M^2/s\ll 1$), the differential cross-section has
the evident factorizable form
\begin{equation}
\frac{d\sigma_{H}}{dz}=\sigma_{b \bar b}(s)\cdot D_{\bar b \to H}(z),
\label{frag}
\end{equation}
where $z=2E_{B_c}/\sqrt{s}$, and $D_{\bar b \to H}(z)$ 
is interpreted as the $\bar b \to H+X$ fragmentation function with $H$
being the $\bar b c$-quarkonium [1-3].

Recently, the functions of fragmentation into the
$S$-, $P$-, and $D$-wave states for the $\bar b c $-quarkonium production 
in $e^+e^-$ collisions have been found [1-4]. 
Everywhere below mentioning the fragmentation model
we will imply expressions like  (\ref{frag}).

The description is more complicated for $\gamma \gamma$ and hadron-hadron
collisions. As for these interactions, in addition to the mechanism of
the $\bar b$ quark production with the following fragmentation into the
$\bar b c$-quarkonium, there is a new type of diagrams, corresponding to 
the photon (gluon) dissociation into the pair of heavy quarks with the
following recombination of the quarks into the $\bar b c$-quarkonium.

In the photon-photon collisions, one can separate three gauge invariant
subgroups (6+6+8) within complete set of 20 diagrams in the
leading  Born approximation, so that subgroups allow one to have
the self-consistent
interpretations: the $\bar b$ quark fragmentation,  $c$ quark
one, and  recombination. The contribution of the first mentioned
type is quite reliably described by  expression (\ref{frag}). For the
$c$ quark fragmentation diagrams one observes a strong deviation from the
picture of the fragmentation model. The recombination contribution dominates
in the complete region of kinematical variables. As was shown in
[5], the given conclusion is valid for the $S$-wave state production 
as well as the $P$-wave one [6].

The process of hadronic production of the $\bar b c$-quarkonium is 
more difficult for the analysis. In the leading order of the 
perturbation theory
it is described by 36 diagrams in the fourth order over $\alpha_s$.
At present, there are results of calculations by several groups
giving controversial numerical values
and conclusions for the $\bar b c$-quarkonium production cross-sections  
[7-10]. Moreover, it was offered to simplify the consideration
of the heavy quarkonium production by  its reduction to the
straightforward fragmentation of $\bar b$ quark with the usage of the
fragmentation functions, derived for $e^+e^-$-annihilation case [11].

In this paper we give the comparative analysis of the results
by different papers devoted to the production of the $S$-wave states.
We prove that the straightforward fragmentation model does not properly
describe
the $\bar b c$-quarkonium production. Next, we present the results of 
calculations for the $P$-wave level production over the complete set of
the leading order diagrams.

\section{Calculation technique}
The $A^{SJj_z}$ amplitude of the $B_c$ meson production can be 
expressed through 
the amplitude of four free quarks production $T^{Ss_z}(p_i,k({\bf q}))$ 
and the orbital wave function of the $B_c$ meson, $\Psi^{Ll_z}({\bf q})$, 
in the meson rest frame as
\begin{equation}
A^{SJj_z}=\int T^{Ss_z}(p_i,k({\bf q}))\cdot 
\left (\Psi^{Ll_z}({\bf q}) \right )^* \cdot
C^{Jj_z}_{s_zl_z} \frac{d^3 {\bf q}}{{(2\pi)}^3},
\label{int}
\end{equation}
where $J$ and $j_z$ are the total spin of the meson and its projection on $z$ 
axis in the $B_c$ rest frame, correspondingly; $L$ and $l_z$ are the orbital
momentum and its projection; $S$ and $s_z$ are the sum of quark spins and its
projection; $C^{Jj_z}_{s_zl_z}$ are the Clebsh-Gordan coefficients;
$p_i$ are four-momenta of $B_c$, $b$ and $\bar c$,
${\bf q}$ is the three-momentum of $\bar b$ quark in the $B_c$ meson rest
frame; $k({\bf q})$ is the four-momentum, obtained from the
four-momentum $(0,{\bf q})$ by the Lorentz transformation from the
$B_c$ rest frame to the system, where the calculation of
$T^{Ss_z}(p_i,k({\bf q}))$ is performed. Then, the four-momenta
of $\bar b$ and $c$ quarks, composing the $B_c$ meson, will be determined 
by the following formulae with the accuracy up to $|{\bf q}|^2$ terms
\begin{equation}
\begin{array}{c}
p_{\bar b}=\frac{m_b}{M}P_{B_c}+k({\bf q}), \\
p_{c}=\frac{m_c}{M}P_{B_c}-k({\bf q}),
\end{array}
\label{mom}
\end{equation}
where $m_b$ and $m_c$ are the quark masses, $M=m_b+m_c$,
and $P_{B_c}$ is the $B_c$ momentum. Let us  note that for the 
$P$-wave states it is
enough to take into account only terms, linear over ${\bf q}$ in 
eq.(\ref{int}), and ${\bf q}=0$ in the $S$-wave production.

The product of spinors $v_{\bar b} \bar u_c$, corresponding to the
$\bar b$ and $c$ quarks in the $T^{Ss_z}(p_i,k({\bf q}))$ amplitude of
eq.(\ref{int}), should be substituted by the projection operator
\begin{equation}
{\cal P} (\Gamma )=\sqrt{M} \left (\frac{\frac{m_b}{M} \hat P_{B_c}
+\hat k-m_b}{2m_b}
\right ) \Gamma \left (\frac{\frac{m_c}{M}\hat P_{B_c}-\hat k+m_c}{2m_c}
\right ),
\end{equation}
where $\Gamma=\gamma^5$ for $S=0$, or $\Gamma=\hat \varepsilon^*(P_{B_c},s_z)$
for $S=1$, where $\varepsilon(P_{B_c},s_z)$ is the polarization vector for the
spin-triplet state.

For the sake of convenience, one can express the ${\cal P}(\Gamma )$ operator
through the spinors of the following form
\begin{equation}
\begin{array}{c}
v_b'(p_b+k,\pm)=\left ( 1-\frac{\hat k}{2m_b} \right )v_b(p_b,\pm),\\
u_c'(p_c-k,\pm)=\left ( 1-\frac{\hat k}{2m_c} \right )u_c(p_c,\pm),
\end{array}
\label{spin}
\end{equation}
where $v_b(p_b,\pm)$ and $u_c(p_c,\pm)$ are the spinors with the given
projection of quark spin on $z$ axis in the $B_c$ meson rest frame.
Note, that the spinors in eq.(\ref{spin}) satisfy the Dirac equation
for the antiquark with the momentum $p_b+k$ and mass $m_b$ or for the
quark with the momentum $p_c-k$ and mass $m_c$ up to the linear order over
$k$ ( i.e. over ${\bf q}$, too), correspondingly.

One can easily show that the following equalities take place
\begin{equation}
\begin{array}{c}
\sqrt{\frac{2M}{2m_b2m_c}}\frac{1}{\sqrt{2}}\{v_b'(p_b+k,+)\bar u_c'(p_c-k,+)-
v_b'(p_b+k,-)\bar u_c'(p_c-k,-)\}=\\
={\cal P}(\gamma^5)+O(k^2),\\ \\

\sqrt{\frac{2M}{2m_b2m_c}}v_b'(p_b+k,+)\bar 
u_c'(p_c-k,-)=\\
={\cal P}(\hat \varepsilon^*(P,-1))+O(k^2),\\ \\

\sqrt{\frac{2M}{2m_b2m_c}}\frac{1}{\sqrt{2}}\{v_b'(p_b+k,+)\bar u_c'(p_c-k,+)+
v_b'(p_b+k,-)\bar u_c'(p_c-k,-)\}=\\
={\cal P}(\hat \varepsilon^*(P,0))+O(k^2),\\ \\

\sqrt{\frac{2M}{2m_b2m_c}}v_b'(p_b+k,-)\bar 
u_c'(p_c-k,+)=\\
={\cal P}(\hat \varepsilon^*(P,+1))+O(k^2).
\end{array}
\end{equation}
In the $B_c$ rest frame, the polarization vectors of the spin-triplet
state have the form
\begin{equation}
\begin{array}{l}
\varepsilon^{rf}(-1)=\frac{1}{\sqrt{2}}(0,1,-i,0), \\
\varepsilon^{rf}(0)=(0,0,0,1), \\
\varepsilon^{rf}(+1)=-\frac{1}{\sqrt{2}}(0,1,i,0).
\end{array}
\end{equation}
In calculations the Dirac representation of $\gamma$-matrices is used and
the following explicit form of the spinors is applied
\begin{equation}
\begin{array}{cc}
u(p,+)={ \frac{1}{\sqrt{E+m}}
\left (\begin{array}{c}
E+m\\
0\\
p_z\\
p_x+ip_y
\end{array}
\right ) },
&
u(p,-)={ \frac{1}{\sqrt{E+m}}
\left (\begin{array}{c}
0\\
E+m\\
p_x-ip_y\\
-p_z
\end{array} 
\right ) } 
\\
&\\
v(p,+)={ -\frac{1}{\sqrt{E+m}}
\left (\begin{array}{c}
p_z\\
p_x+ip_y\\
0\\
E+m
\end{array}
\right ) },
&
v(p,-)={ \frac{1}{\sqrt{E+m}}
\left (\begin{array}{c}
p_x-ip_y\\
p_z\\
0\\
E+m
\end{array}
\right ) }
\end{array} 
\end{equation}
The $S$-wave production amplitude can be written down as
\begin{equation}
A^{Ss_z}=iR_S(0)\sqrt{\frac{2M}{2m_b2m_c}}
\sqrt{\frac{1}{4\pi}}
\left (T^{Ss_z}\left (p_i,k({\bf q}=0)\right )\right ),
\label{main0}
\end{equation}
where $R_S(0)$ is the radial wave function at the origin, so that
$$
R_S(0) = \sqrt{\frac{\pi}{3}}\; \tilde f_{B_c},
$$
and the $\tilde f_{B_c}$ value is related with the leptonic constants
of pseudoscalar and vector $B_c$ states
\begin{eqnarray}
\langle 0| J_\mu(0)|V\rangle & = & i f_V M_V\; \epsilon_\mu\;, \nonumber\\
\langle 0| J_{5\mu}(0)|P\rangle & = & i f_P p_\mu\;, \nonumber
\end{eqnarray}
where $J_{\mu}(x)$ and $J_{5\mu}(x)$ are the vector and axial-vector currents
of the constituent quarks. Then the account for hard gluon corrections
in the first order over $\alpha_s$ [12] results in
\begin{eqnarray}
\tilde f & = & f_V\; \bigg[1 - \frac{\alpha_s^H}{\pi}
\biggl(\frac{m_2-m_1}{m_2+m_1}\ln\frac{m_2}{m_1} -\frac{8}{3}\biggr)\bigg]\;,\\
\tilde f & = & f_P\; \bigg[1 - \frac{\alpha_s^H}{\pi}
\biggl(\frac{m_2-m_1}{m_2+m_1}\ln\frac{m_2}{m_1} - 2\biggr)\bigg]\;,
\end{eqnarray}
where $m_{1,2}$ are the masses of quarks, composing the quarkonium. For the
vector currents of quarks with equal masses, the BLM procedure of the scale 
fixing in the "running" coupling constant of QCD [13] gives (see paper by 
M.B.Voloshin in ref.[12])
$$
\alpha_s^H = \alpha_s^{\overline{\rm MS}}(e^{-11/12}m_Q^2)\;.
$$
The estimates of the $\tilde f_{B_c}$ value within the potential models
have the essential uncertainty, $\tilde f_{B_c}=500 \pm 100$ MeV [14].
The QCD sum rule estimate of the $f_{B_c}$ value for the pseudoscalar state
gives $f_{B_c}= 385\pm 25$ MeV [15], which is in a good agreement with
the evaluation in the framework of recent lattice computations [16],
where $f_{B_c}=395(2)$ MeV with the error bar, giving the statistical
uncertainty only. The $\tilde f_{B_c}$ estimate strongly depends on
the $\alpha_s^H$ scale choice, which is not yet calculated in the BLM
procedure. So, we use $\tilde f_{B_c}= 570$ MeV.

For the $P$-wave states in eq.(\ref{int}), the
$T^{Ss_z}\left (p_i,k({\bf q})\right )$ amplitude can be expanded
into the Taylor series up to the terms linear over ${\bf q}$.
Then one gets
\begin{equation}
A^{SJj_z}=iR_P'(0)\sqrt{\frac{2M}{2m_b2m_c}}
\sqrt{\frac{3}{4\pi}}C^{Jj_z}_{s_zl_z}
{\cal L}^{l_z}\left (T^{Ss_z}\left (p_i,k({\bf q})\right )\right ),
\label{main}
\end{equation}
where $R_P'(0)$ is the first derivative of the radial wave function at the 
origin, and ${\cal L}^{l_z}$ has the following form
\begin{equation}
\begin{array}{l}
{\cal L}^{-1}=\frac{1}{\sqrt{2}}\left (\frac{\partial}{\partial q_x}
+i\frac{\partial}{\partial q_y} \right ), \\
{\cal L}^0=\frac{\partial}{\partial q_z}, \\
{\cal L}^{+1}=-\frac{1}{\sqrt{2}}\left (\frac{\partial}{\partial q_x}
-i\frac{\partial}{\partial q_y} \right ), 
\end{array}
\label{dif}
\end{equation}
where $\frac{\partial}{\partial q_x}$,
$\frac{\partial}{\partial q_y}$, $\frac{\partial}{\partial q_z}$ are the
differential operators acting on $T^{Ss_z}\left (p_i,k({\bf q})\right )$ as
the function of ${\bf q}=(q_x,q_y,q_z)$ at ${\bf q}=0$.

As all considered matrix elements are calculated in the system 
distinct from the $B_c$ rest frame, the four-momentum $k({\bf q})$
has been calculated by the following formulae
\begin{equation}
\begin{array}{l}
k^0=\frac{{\bf  v} \cdot {\bf q}}{\sqrt{1- {\bf v}^2}}, \\
{\bf k}={\bf q} +(\frac{1}{\sqrt{1- {\bf v}^2}}-1)\frac{{\bf  v} 
\cdot {\bf q}}
{{\bf v}^2}{\bf  v},
\end{array}
\label{Lor}
\end{equation}
where ${\bf  v}$ is the $B_c$ velocity in the system, where the calculations
are performed. The matrix element
$T^{Ss_z}\left (p_i,k({\bf q})\right )$ is computed, so that the
four-momenta of $\bar b$ and $c$ quarks are determined by eq.(\ref{mom}),
taking into account eq.(\ref{Lor}).

The first derivatives in eq.(\ref{dif}) are substituted by the 
following approximations
\begin{equation}
\frac{\partial T^{Ss_z}\left (p_i,k({\bf q})\right )}{\partial q_j}
\vert_{{\bf q=0}}\approx \frac{T^{Ss_z}\left (p_i,k({\bf q}^j)\right )
-T^{Ss_z}\left (p_i,0\right )}{\triangle},
\end{equation}
where $\triangle$ is  some small value, and ${\bf q}^j$ have the following form
\begin{equation}
\begin{array}{c}
{\bf q}^x=(\triangle,0,0),\\
{\bf q}^y=(0,\triangle,0),\\
{\bf q}^z=(0,0,\triangle).
\end{array}
\end{equation}
With the chosen values of quark masses and interaction energies, the 
increment value $\triangle=10^{-5}$ GeV has provided the stability of
4-5 meaning digits in the  squared matrix elements summed over $j_z$ 
for all $P$-wave states with the given value of $J$ and $S$,
when one has performed the Lorentz transformations along the beam axis or
the rotation around the same axis.

One has to note that because of such transformations, 
the new vectors $k({\bf q}^j)$
do not correspond to the transformed old vectors. Therefore, the applied test
is not only a check of the correct typing of the
$T^{Ss_z}\left (p_i,k({\bf q})\right )$ amplitude, but it is also the check
of correct choice of the phases in eq.(\ref{main}).

The  matrix element $A^{SJj_z}$ squared,  which is calculated by the method
described above, must be summed over $j_z$ as well as the spin states of free
$b$ and $\bar c$ quarks. It also must be  averaged over spin projections
of initial particles.

The phase space integration has been made by the Monte Carlo method of RAMBO
program [11].

\section{$S$-wave states}
Let us consider the calculation results for the $gg$-production of the
$S$-wave $\bar b c$-quarkonium levels ($B_c$ and $B_c^*$) in
comparison with the values given in other papers. The cross-section of 
the process under consideration is proportional to the 
$\tilde f_{B_c}$ squared 
as well as to the fourth power of $\alpha_s$. To compare the results of
different papers, we rescale all the numbers to the values, determined
by the same set of $\alpha_s=0.2$ and $\tilde f_{B_c}=570$ MeV, which are
used in the given calculations. We also fix the mass values of
$m_b=4.8$ GeV and $m_c=1.5$ GeV.

The cross-sections of the gluonic production of $B_c^{(*)}$ versus
the total energy of the $gg$ collisions are presented in Fig.~1.
In addition to the results of the complete numerical calculations of the
$O(\alpha_s^4)$ contribution, we give also the values, obtained
in the fragmentation model and, hence, calculated as the product of the
$gg$-production cross-section for the $b \bar b$-pair and the 
probability of the $\bar b \to B_c^{(*)}$ fragmentation, calculated at
the same set of parameters.

As one can see in this figure, there is a good agreement of our previous
calculations [7] with the results of refs.[8,10]. The values, given in 
two other papers of ref.[9], are approximately 
three times greater than our results. 
In this paper we have recalculated the cross-sections in the
axial gauge, used in ref.[9], in contrast to the covariant Feynman gauge
applied in our previous consideration [7]. The corresponding numerical
results have not changed after the replacement of the gauge. This fact
points to errors in the computations, performed in ref.[9].

In the studied region of energies, the ratio of the $B_c$ and $B_c^*$ 
yields, $R=\sigma_{B_c^*}/\sigma_{B_c}\simeq 3$, strongly
deviates from $R\simeq 1.4$ predicted by the fragmentation model. One
can also see in Fig.~1, that in the region of the applicability determined by
the condition $M^2/s\ll 1$, the fragmentation model gives the total 
cross-section, which is essentially smaller than the exact result evaluated
over the complete set of diagrams.

As one can see in Fig.~2, the total cross-section of the $B_c^{(*)}$
production is basically accumulated in the region of the transverse momenta
close to $p_T \sim M_{B_c}$. One could expect, that the fragmentation 
mechanism begins to dominate at large $p_T$. Indeed, the $p_T$-distribution
shown in Fig.~2 for $B_c$ and $B_c^*$ at $\sqrt{\hat s} =100$ GeV, points to 
the
fact, that there is a quite good coincidence of the distribution tails
obtained from exact perturbative calculations and in the
fragmentation model. Numerically, this agreement takes place at
$p_t> 40$ GeV for $B_c$. Thus, one can draw the certain conclusion, that
the fragmentation contribution is not dominant in the $gg$-collisions.
It also does not describe  the distribution over the variable
defined as $z=2|\vec P_{B_c}|/\sqrt{\hat s}$. This fact is evident in
Fig.~3. At  $z$ values close to unit, the fragmentation model overestimates
the exact perturbative result.

To get the cross-section and $p_T$ spectra in hadron-hadron collisions,
one should convert the partonic $gg$ cross-section with the
distribution functions of gluons in the initial hadrons. We use the
parametrization of ref.[17] for the parton distributions at the fixed
virtuality scale $Q=2m_b\sim 10$ GeV. 
In the framework of the given approximation,
we fix also the coupling constant value $\alpha_s=0.2$.

The convolution result is presented in Fig.~4 for the energy of the FNAL
Tevatron ($\sqrt{s}=1.8$ TeV). The histograms correspond to the
exact perturbative results, whereas the curves are calculated according to
the following formula
\begin{eqnarray}
\frac{d\sigma}{dp_T}&(&\bar p p\rightarrow H_{(p_T)}x)=
\sum_{i,j}\int dx_1dx_2dz f_{i/p}
(x_1,\mu)f_{j/\bar p}(x_2,\mu)\times\nonumber \\
\times \frac{d\hat\sigma}{dp_T}
&(&ij\rightarrow \bar b(p_T/z)+x)\times D_{\bar b\rightarrow H}(z,\mu),
\label{strfun}
\end{eqnarray}
where $D(z,\mu)$ is the function of the $\bar b\rightarrow H$ fragmentation
with  $H=B_c, B^*_c...,\, d\hat\sigma/dp_T$ is the differential
cross-section of the $\bar b$ quark production, and
$f_{i/A}(x,\mu)$ is the distribution of the $i$-kind parton in the $A$-hadron.

As one can see in the figure, the curves certainly deviate from the
histograms of the exact perturbative calculations in the all region of
$p_T$. At small $p_T$, the fragmentation model gives the overestimation,
whereas at large $p_T$, contrary, it underestimates the exact result.
It is significant to note the fact, that the perturbative calculations
and the model give the different values for the ratio of the differential 
cross-sections of $B_c^*$ and $B_c$. One finds the yield ratio 
$R \sim 2\div 3$, whereas the model gives $R \sim 1.3\div 1.5$.

The qualitative agreement between the model and perturbative
calculations for the $d\sigma/dp_T$ distribution of 
$B_c$ meson has allowed the authors of ref.[8] to conclude on the satisfactory
description of the exact perturbative $O(\alpha_s^4)$-contribution by
the fragmentation model. The misleading is related to the fact, that those
values of the $gg$-collision energy are included in the integration region of
expression (\ref{strfun}), where the approximation of the fragmentation does
not work or it is not strictly defined, i.e., where the $M^2/s \ll 1$ 
condition is evidently not valid. As one can see in Fig.~1 at these low
energies, the cross-section in the fragmentation model and that of the
perturbative result have the different dependency on the energy. The latter
fact is because of the use of the two-particle phase space in the fragmentation
approximation of the heavy quark production instead of the real three-particle
phase space ($B_c$, $\bar b$ and $c$ are in the final state).

To avoid the uncertainties related with the low energy of $gg$-collisions,
we have started the integration over $\hat s$ in expression (\ref{strfun}) 
from $\sqrt{\hat s} > 60$ GeV. As one can see in Fig.~5, the given cut
drastically changes the relation between the fragmentation model
contribution and the exact perturbative result. It indicates
that the fragmentation model can not provide an adequate approximation for
the correct description of the hadronic $B_c$ meson production.

One has to note, that the given conclusion does not depend on the definite
choice of the structure functions for the initial hadrons as well as on
a special modification of the fragmentation expression.

Thus, the conclusion drawn in ref.[8] about the dominance of the fragmentation
mechanism in the hadronic production of $B_c$ mesons, is  incorrect.

\section{Production of $P$-wave states}

As it was mentioned above, to get the cross-section for the production of
the $P$-wave quarkonium levels it is 
necessary to calculate the first derivative
of the quark production matrix element over $\vec q$,  the relative
momentum of the quarks inside the $B_c$ meson. This procedure of calculations
has been tested by comparison of the numerical computation of the
fragmentation functions for the $\bar b$ quark into the $P$-wave
states with known analytical expressions for the corresponding
functions obtained in ref.[4]. We have found a good agreement between the
$^1P_1$, $^3P_0$, $^3P_1$, and $^3P_2$-level cross-sections calculated
numerically, and those  obtained from the analytical expressions
 for the same set of parameters,
$m_b$, $m_c$, $|R_P'(0)|^2$ and $\alpha_s$. The total cross-sections of the
gluonic production of $P$-wave levels are presented versus the total energy
of the collisions in Tab.~1. The dependence of the cross-section, 
summed over the spin states
of the $P$-levels,  on the energy of the interacting gluons is shown in Fig.~6. 
This dependence can be approximately described by the following expression
\begin{equation}
\sigma_{B_c(L=1)}=25.\cdot \left ( 1- \left ( \frac{2(m_b+m_c)}{\sqrt{s}}
\right ) \right )^{1.95} 
\cdot \left ( \frac{2(m_b+m_c)}{\sqrt{s}} \right )^{1.2} {\rm pb}.
\label{emp}
\end{equation}
The contribution, determined by the fragmentation model, is also shown in the
same figure, where it is found as the corresponding total $gg\to b \bar b$ 
cross-section multiplied by the integral probability of the
$b \to B_c(L=1)$ fragmentation, $W=5.34\cdot 10^{-5}$. The $c$-quark
fragmentation is suppressed by the order of magnitude in the hadron collisions,
and we will neglect it below.

As one can see in Fig.~6, the total cross-section of the $P$-wave level
production is much greater than the value predicted by the fragmentation model,
which has a strict meaning in the region of high energies, and, thereby, the
contribution of the recombination diagrams dominates.   The fragmentation
contribution also does not describe the $d\sigma/dz$ distributions,
which are the differential cross-sections integrated over the transverse 
momentum of
$P$-wave levels. These distributions are shown in Fig.~7. It is interesting to
notice that in contrast to the $\gamma\gamma$-collisions, the result of the
fragmentation model essentially overestimates the exact perturbative
values in the gluonic collisions. This fact, related with the interference of
different contributions, will be discussed elsewhere.

The distributions over the transverse momentum of the produced $P$-wave states
are presented in Fig.~8 in comparison with the results of the
fragmentation model. One can see, that the fragmentation is valid only
on the tails. This fact is quite expected, since the same picture
has been observed in the production of $S$-levels.

Thus, our calculations of the $P$-wave level production cross-sections show
that the total hadronic cross-sections are about 10\%  of the
corresponding cross-sections of the $S$-wave levels, and the recombination
mechanism dominates for the both $S$- and $P$-levels.

The differential distributions for the $P$-wave states calculated by 
convoluting the partonic cross-sections with the gluon
distributions, as it was described in previous Section, are shown in Fig.~9
for the FNAL Tevatron energy, $\sqrt{s}=1.8$ TeV.

As it comes for the $S$-wave states, the calculation over formula (\ref{frag}) 
of the fragmentation model gives a rather approximate qualitative value for the
$P$-level cross-section. The reason of such "description" is generally related
to the uncertainty of the fragmentation approach in the region of
energies close to the kinematical threshold of the reaction, where
the fragmentation model overestimates the cross-section. This incorrect
contribution compensates the underestimation in the region of large energies
and small or moderate $p_T$, where the recombination dominates. This 
results in the approximate description of the exact perturbative distribution
by the fragmentation formula at $p_T\sim 15$ GeV (see Fig.~9). As it was
shown in previous Section, the $\hat s $ cut  clarifies the problem,
and one evidently conclude, that the exact perturbative calculations and the
fragmentation model strongly differ in the predictions.

\section*{Conclusion}
Let us analyse the results of the numerical perturbative calculations
performed for the hadronic production cross-sections of the
$S$- and $P$-wave levels of the $\bar b c$-quarkonium in the leading
$O(\alpha_s^4)$ order of the perturbative QCD.

There are two scales of virtualities in the problem. The first one is of the
order of the heavy quark masses, and it appears in the heavy quark production.
The second scale is determined by the characteristic relative momentum of the
quarks inside the $\bar b c$-meson. Since the latter is much less than the
masses of the produced quarks, the process of the $B_c$ production can
be separated into two steps:

1) the heavy quark production, described in the perturbative QCD, and

2) the forming of the bound state, described by the quarkonium wave function
at the origin.

\noindent
The results, hence, linearly depend on $\alpha_s^4$, ${\tilde f_{B_c}}^2$
and $|R'_P(0)|^2$. Moreover, there are the additional parameters
defining the quark masses, $m_b$ and $m_c$. The bulk of the 
uncertainty of the
results is connected with the $\alpha_s$, ${\tilde f_{B_c}}$,
and $|R'_P(0)|$ values. For instance, the use of the running coupling constant
$\alpha_s(\hat s)$  instead of $\alpha_s(4m_b^2)$ as well as the use of a
smaller value of $\tilde f_{B_c}$ can decrease the cross-section by order
of magnitude.

From our point of view, the use of $\alpha_s(\hat s)$ is not correct, since
the analysis shows that typical virtualities in the production of four heavy
quarks are less than $\hat s$ at large $\hat s$. 
This is why we have fixed the value
of $\alpha_s=0.2$. The comparison with the results of other papers shows, that
we agree with the calculations of the $S$-level production in ref.[8],
but, in contrast to [8], we compare the yields of pseudoscalar and
vector states of $B_c$ in the hadronic production to emphasize the
invalidity of the fragmentation model to the problem under consideration. The
results of ref.[9] disagree with the values given in our paper for both
total cross-section of the pseudoscalar state and the dependence on
the total energy. The production of the vector state was not considered
in ref.[9]. So, our computations, performed in covariant and axial gauges,
point to errors in papers of ref.[9]. 

The exact perturbative cross-sections for the $P$-wave states are found for
the first time. As well as for the hadronic production of $S$-levels, one 
observes
the strong discrepancy in the values of the relative yields of
$^1P_1$, $^3P_0$, $^3P_1$ and $^3P_2$-levels in comparison with the
production in $e^+e^-$-annihilation. This fact points to the different
dominating mechanisms. The fragmentation of the $\bar b$ quark into
$B_c^{(*)}$ is the basic mechanism in $e^+e^-$-annihilation, whereas
the recombination of heavy quarks  dominates in the hadronic
production of $B_c^{(*)}$.

\section*{Acknowledgements}
A.K.~Likhoded thanks E.~Eichten and C.~Quigg for fruitfull
discussions and stimulating remarks.

This work is supported, in part, by the Russian Foundation of Fundametal
Researches. The work of A.V.~Berezhnoy
has been made possible by a fellowship of INTAS Grant 93-2492 and  one
of International Soros Science Education Program Grant A1377 and it
is carried out within the research program of International Center for 
Fundamental Physics in Moscow.
\newpage
\section*{References}
\begin{itemize}
\item[1.]
  L~Clavelli, Phys. Rev. D26 (1982) 1610;\\
  C.-R.~Ji and R.~Amiri, Phys. Rev. D35 (1987) 3318;\\
  C.-H.~Chang and Y.-Q.~Chen, Phys. Lett. B284 (1992) 127.
\item[2.]
E.~Braaten, K.~Cheung, T.C.~Yuan, Phys. Rev. D48 (1993) 4230.
\item[3.]
V.V.~Kiselev, A.K.~Likhoded, M.V.~Shevlyagin, Z. Phys. C63 (1994) 77.
\item[4.]
T.C.~Yuan, Phys. Rev. D50 (1994) 5664;\\
K.~Cheung, T.C.~Yuan, Preprint UCD-95-24, 1995.
\item[5.]
A.V.~Berezhnoy, A.K.~Likhoded, M.V.~Shevlyagin,  Phys. Lett. B342 (1995) 351;\\
K.~Ko\l odzej, A.~Leike, R.~R\"uckl, Phys. Lett. B348 (1995) 219.
\item[6.]
A.V.~Berezhnoy, V.V.~Kiselev, A.K.~Likhoded, Preprint IHEP 95-119,
Protvino, 1995 [hep-ph/9510238].
\item[7.]
A.V.~Berezhnoy, A.K.~Likhoded, M.V.~Shevlyagin, Yad. Fiz. 58 (1995) 730;\\
A.V.~Berezhnoy, A.K.~Likhoded, O.P.~Yushchenko,  Preprint IHEP 95-59, 
Protvino, 1995 [hep-ph/ 9504302].
\item[8.]
K.~Ko\l odzej, A.~Leike, R.~R\"uckl, Phys. Lett. B355 (1995) 337.
\item[9.]
M.~Masetti, F.~Sartogo, Phys. Lett. B357 (1995) 659;\\
S.R.~Slabospitsky, Preprint IHEP 94-53, Protvino, 1994 [hep-ph/9404346].
\item[10.]
C.-H.~Chang, Y.-Q.~Chen, G.-P.~Han and H.-Q.~Jiang, Phys. Lett. B364 (1995) 78.
\item[11.]
K.~Cheung, T.C.~Yuan, Preprint NUHEP-TH-94-20, 1994.
\item[12.]
M.B.~Voloshin, M.A.~Shifman, Sov. J. Nucl. Phys. 47 (1988) 511;\\
E.~Braaten, S.~Fleming, Phys. Rev. D52 (1995) 181;\\
M.B.~Voloshin, Int. J. Mod. Phys. A10 (1995) 2865;\\
V.V.~Kiselev, Preprint IHEP 95-63, Protvino, 1995 [hep-ph/9504313], 
to appear in Int. J. Mod. Phys. A.
\item[13.]
S.J.~Brodsky, G.P.~Lepage, P.B.~Mackenzie, Phys. Rev. D28 (1983) 228.
\item[14.]
S.S.~Gershtein et al., Uspekhi Fiz. Nauk 165 (1995) 3, 
Phys. Rev. D51 (1995) 3613;\\
E.~Eichten and C.~Quigg, Phys. Rev. D49 (1994) 5845.
\item[15.]
S.~Narison, Phys. Lett. B210 (1988) 238;\\
V.V.~Kiselev, A.V.~Tkabladze, Sov. J. Nucl. Phys. 50 (1989) 1063);\\
T.M.~Aliev, O.~Yilmaz, Nuovo Cimento 105A (1992) 827;\\
P.~Colangelo, G.~Nardulli, N.~Paver, Z. Phys. C57 (1993) 43;\\
C.A.~Dominguez, K.~Schilcher, Y.L.~Wu, Phys. Lett. B298 (1993) 190;\\
S.~Reinshagen, R.~R\" uckl, Preprints CERN-TH.6879/93 and MPI-Ph/93-88,
1993;\\
E.~Bagan et al., Z. Phys. C64 (1994) 57;\\
M.~Chabab, Phys. Lett. B325 (1994) 205.
\item[16.]
S.~Kim, Preprint SNU-TP-95-088, hep-lat/9511010, 1995.
\item[17.] 
J.Botts et al., CTEQ Coll., Preprint ISU-NP-92-17, MSUHEP-92-27 (1992).
\end{itemize}

\newpage
\section*{Figure captions}
\begin{itemize}
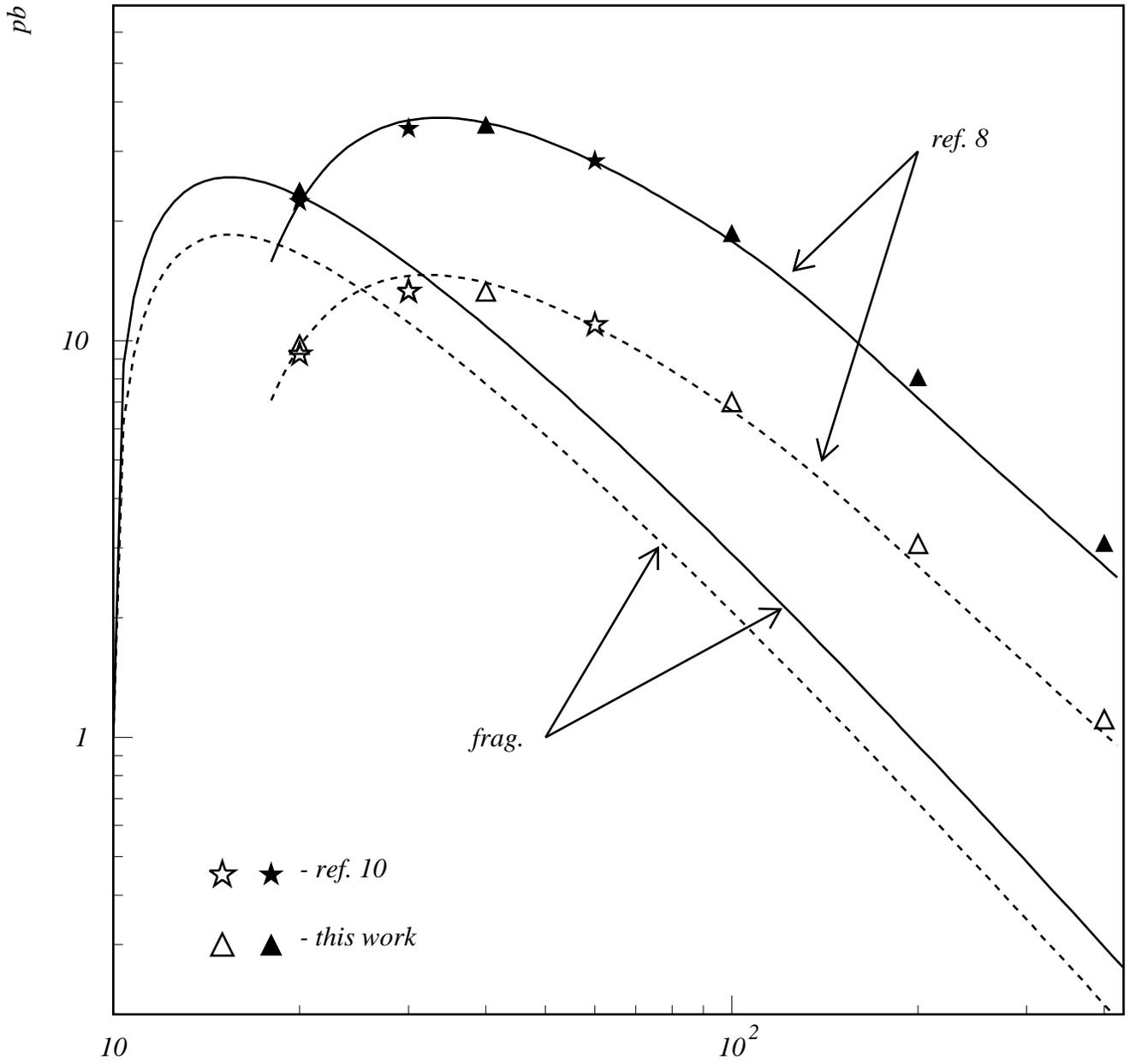
\item[Fig. 1.] The total cross-section of the gluon-gluon production of
$B_c$ (empty triangle) and $B_c^*$ (solid triangle) in comparison with the
predictions of the fragmentation model (frag.) and the results of ref.[8]
(curves) and ref.[10] (stars).
\item[Fig. 2.] $d\sigma/dp_T$ at $\sqrt{\hat s}=100$ GeV for $B_c^*$ (solid
histogram) and $B_c$ (dashed histogram) in comparison with the
predictions of the fragmentation model  (solid and dashed curves for
$B_c^*$ and $B_c$, respectively), and the result of ref.[8] for $B_c$
(dots).
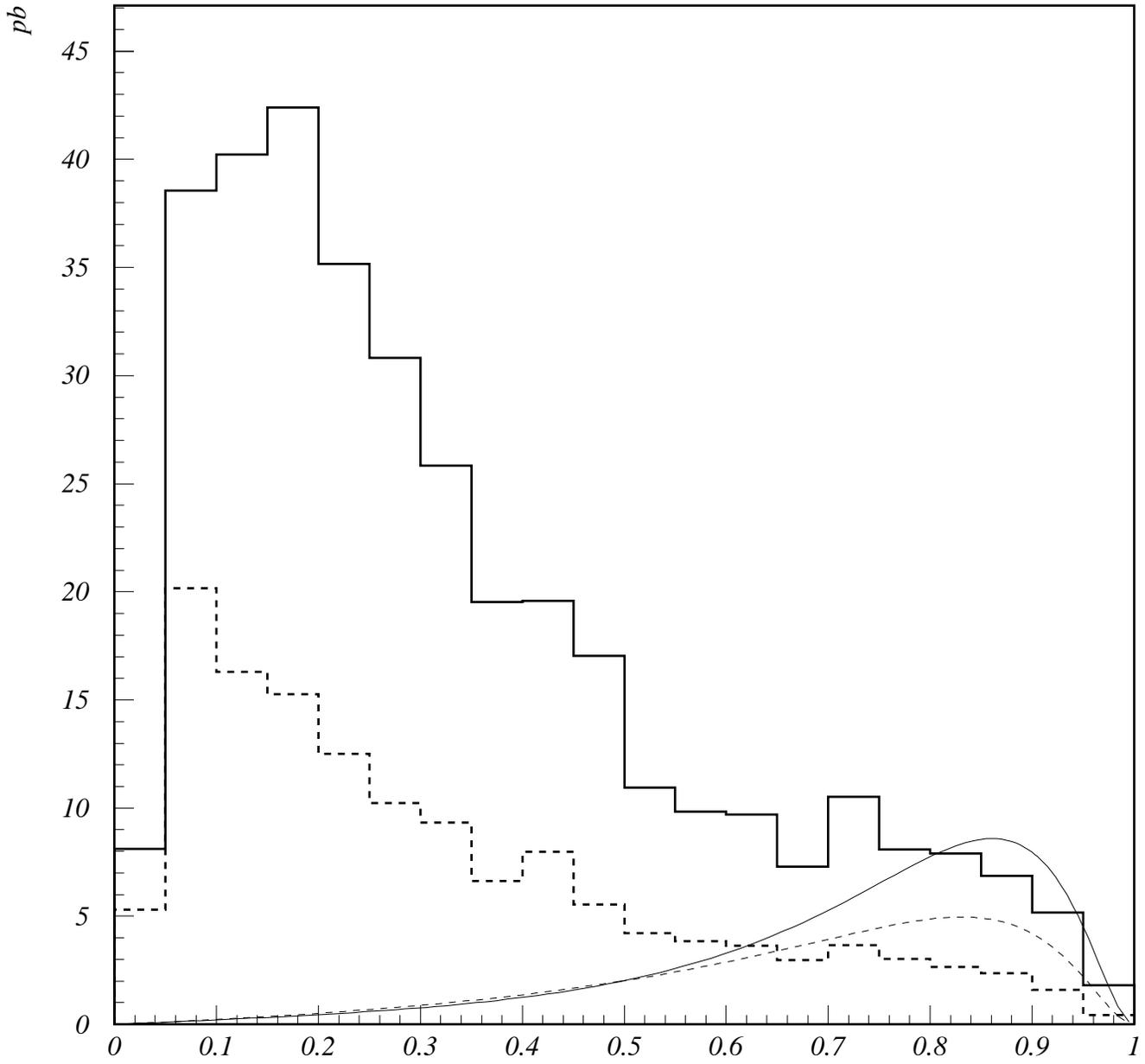
\item[Fig. 3.] $d\sigma/dz$ at $\sqrt{\hat s}=100$ GeV for $B_c^*$ (solid
histogram) and $B_c$ (dashed histogram) in comparison with the
predictions of the fragmentation model (curves).
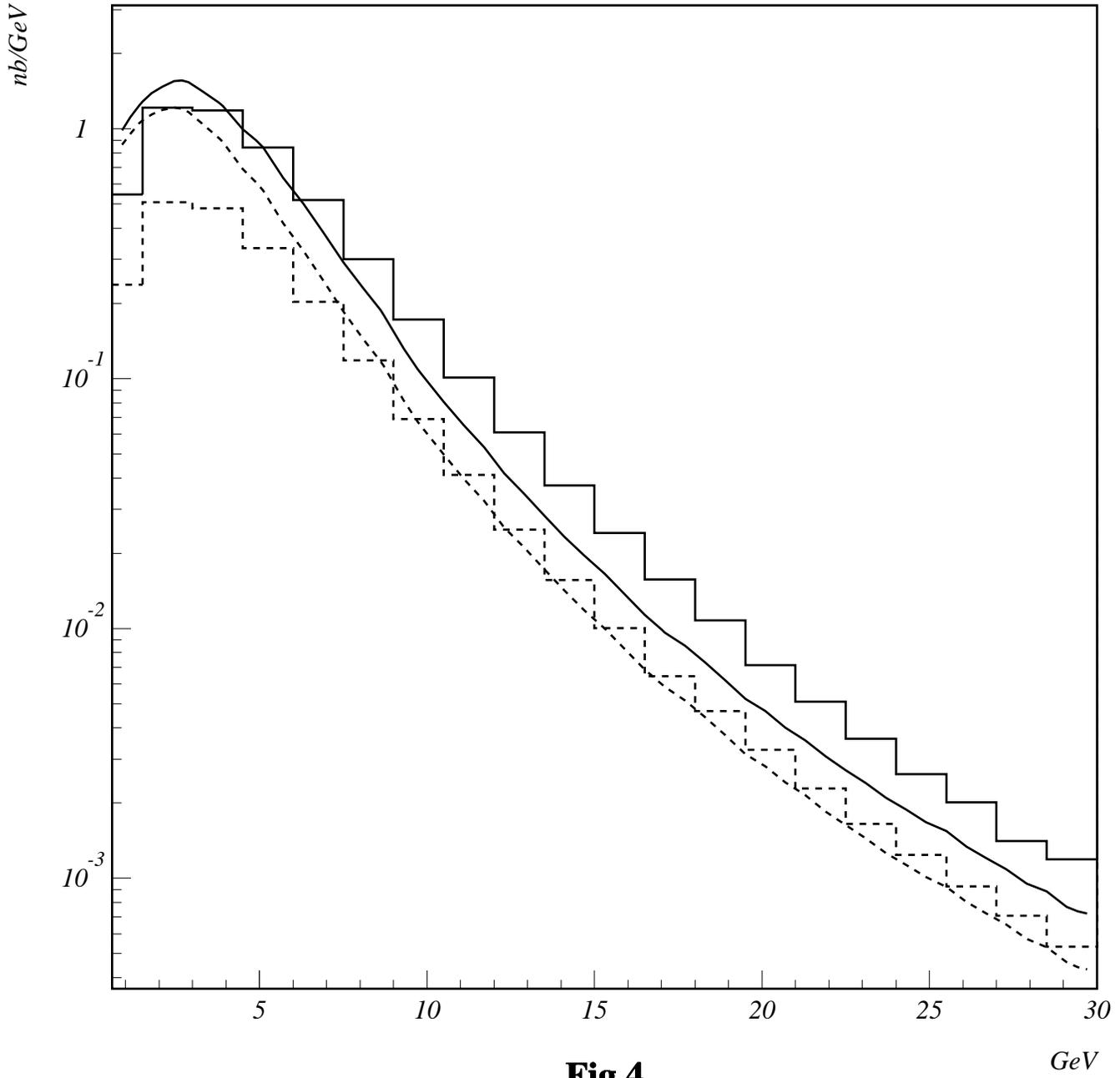
\item[Fig. 4.] $d\sigma/dp_T$ in the hadronic production of $B_c^*$ (solid
histogram) and $B_c$ (dashed histogram) in comparison with the
predictions of the fragmentation model (curves) 
at the $p\bar p$-collision energy $\sqrt{s}=1.8$ TeV.
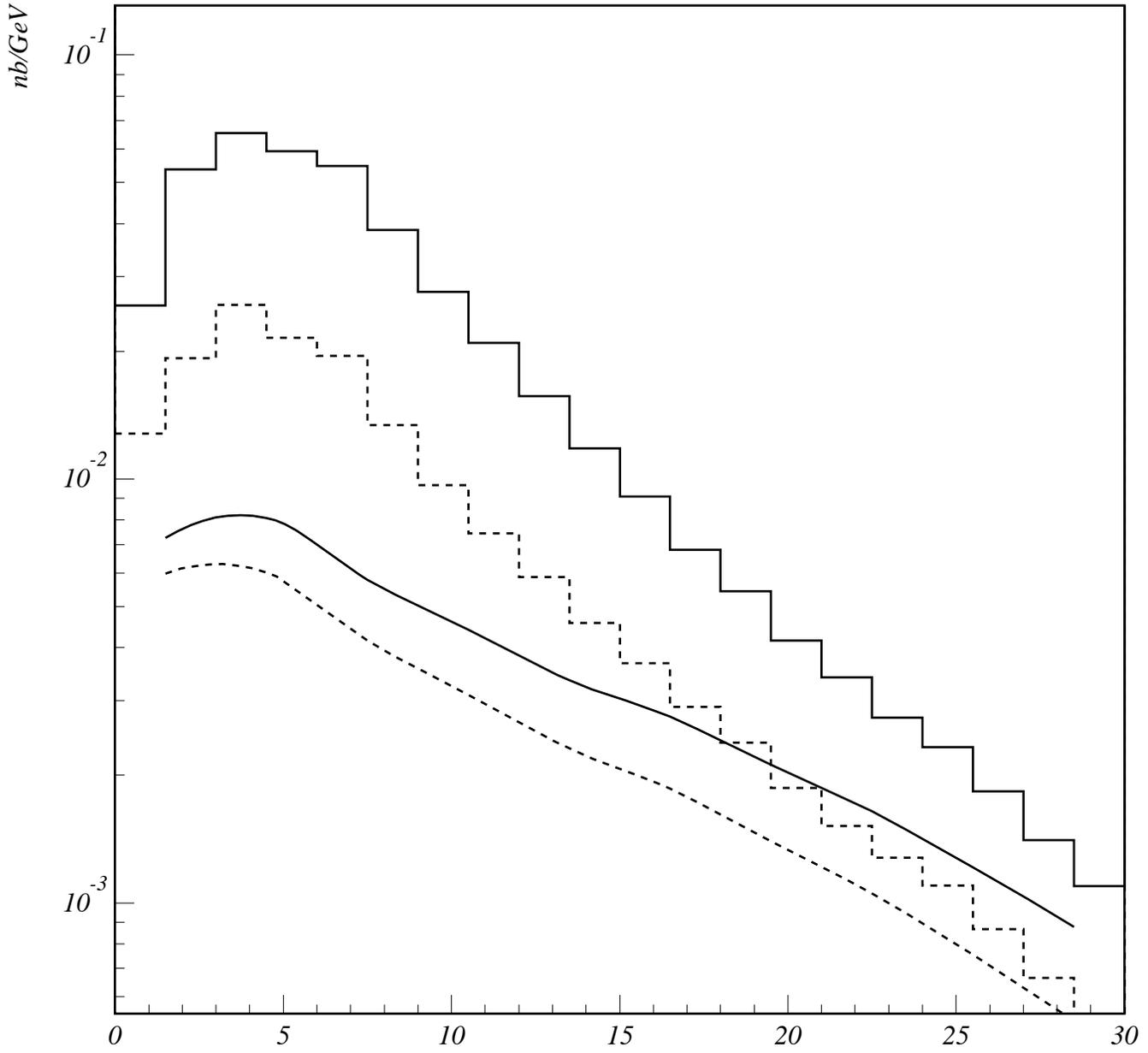
\item[Fig. 5.] The same as in Fig.~4 with the $\sqrt{\hat s}> 60$ GeV cut off.
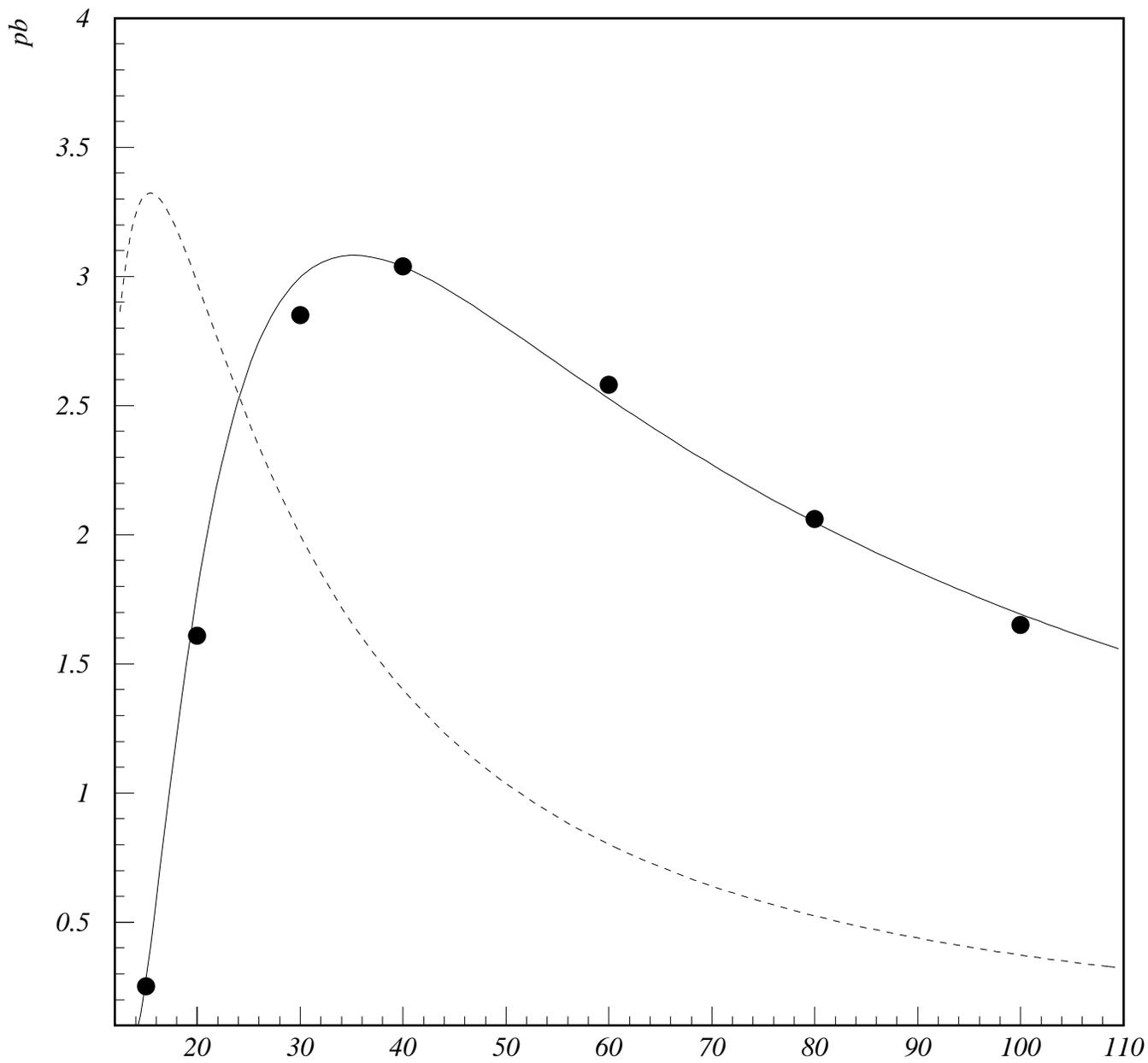
\item[Fig. 6.] The total cross-section summed over spins of the $P$-levels 
in the gluon-gluon production (dots) and its fit (solid line curve) in
comparison with the prediction of the fragmentation model
(dashed line curve).
\item[Fig. 7.] $d\sigma/dz$ at $\sqrt{\hat s}=100$ GeV for $^1P_1$-,
$^3P_0$-, $^3P_1$-, and $^3P_2$-levels (a, b, c and d figures, correspondingly)
in comparison with the predictions of the fragmentation model (curves).
\item[Fig. 8.] $d\sigma/dp_T$ with the same notations as in Fig.~7.
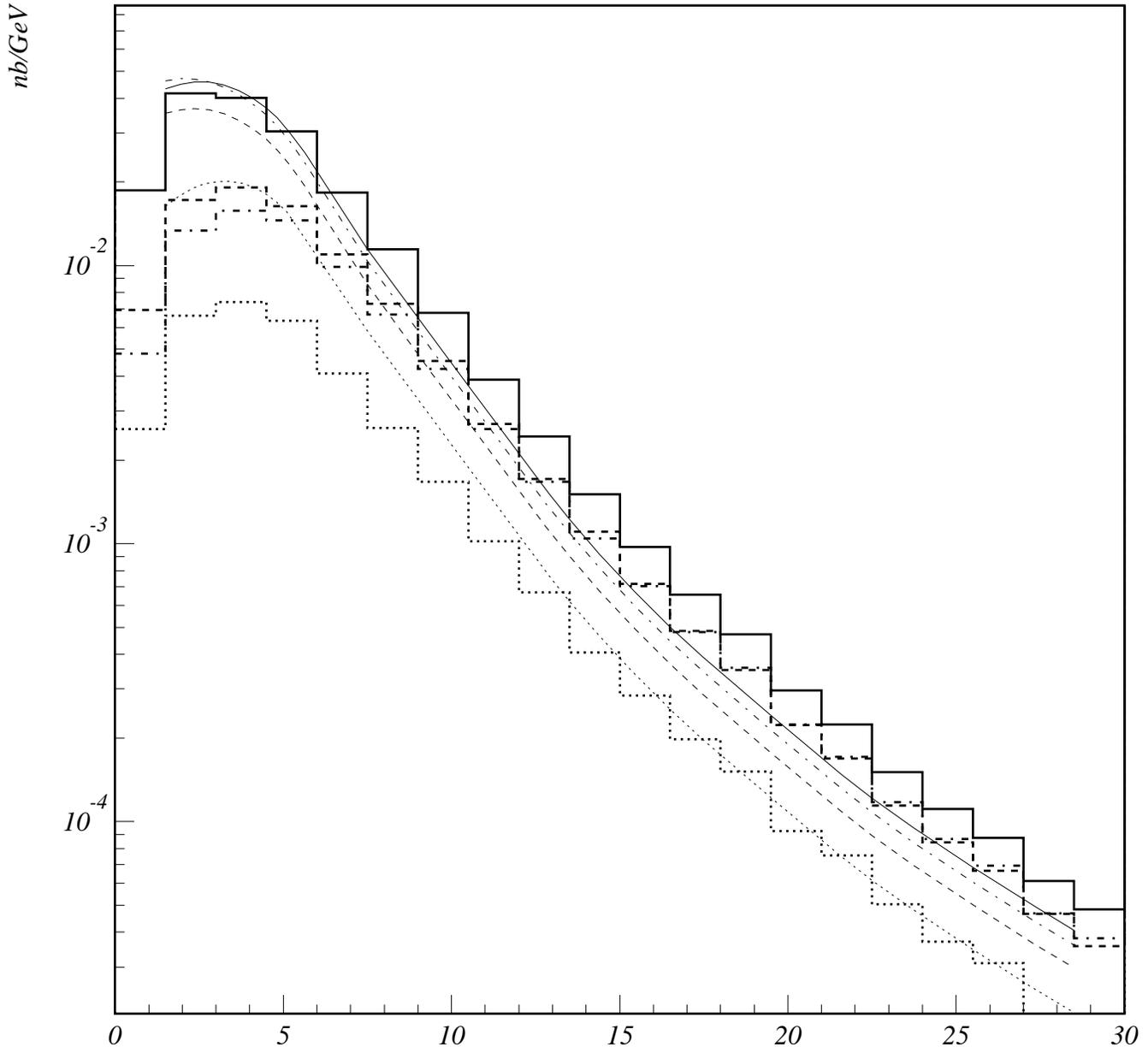
\item[Fig. 9.] The differential cross-section of the hadronic $p\bar p$
production of the $P$-level $B_c$ mesons at $\sqrt{s}=1.8$ TeV
(histograms) in comparison with the fragmentation model results:
$^3P_2$-- solid curve, $^3P_0$-- dotted curve, $^3P_1$-- dash-dotted curve,
and $^1P_1$-- dashed curve.
\end{itemize}
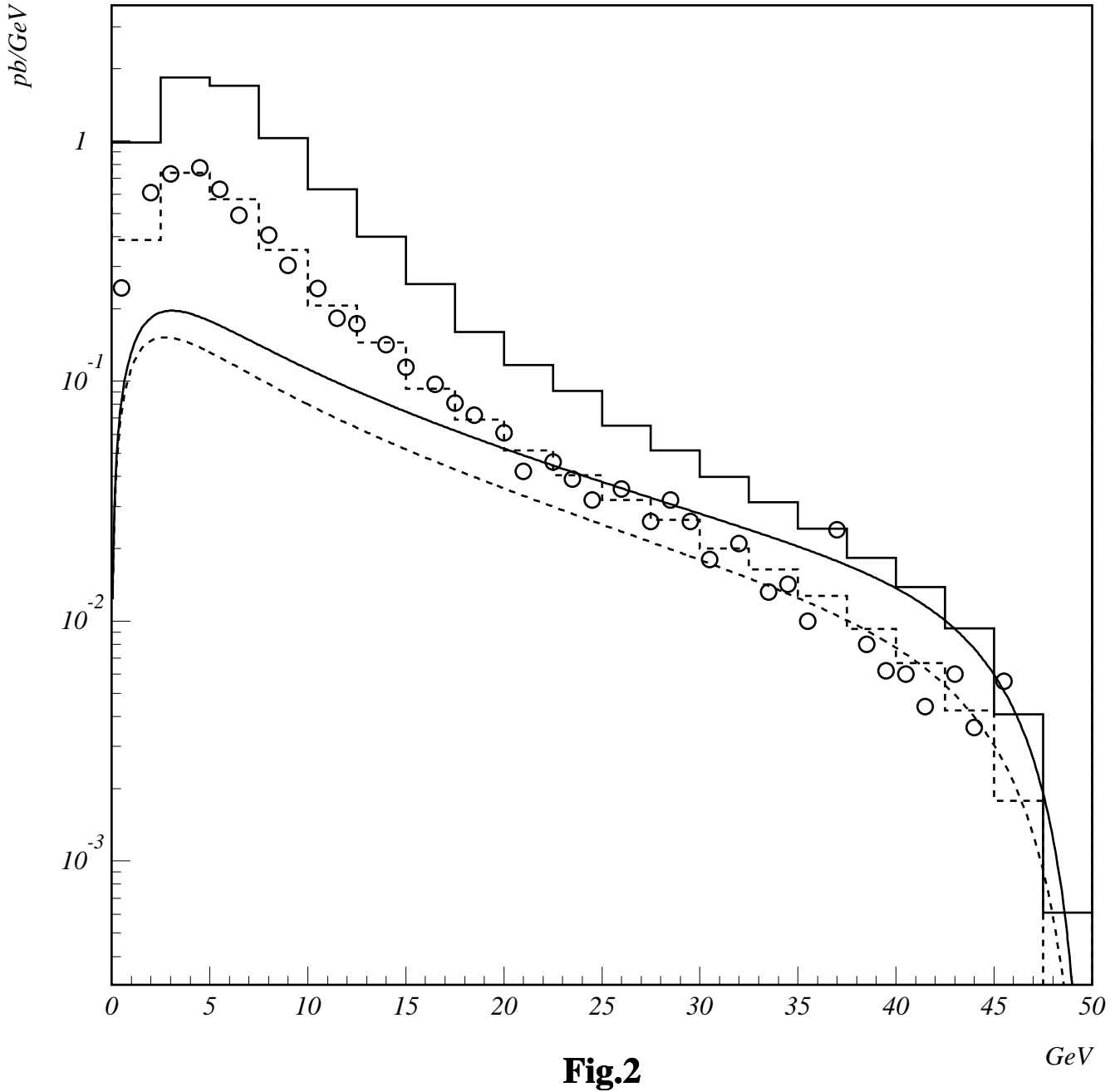
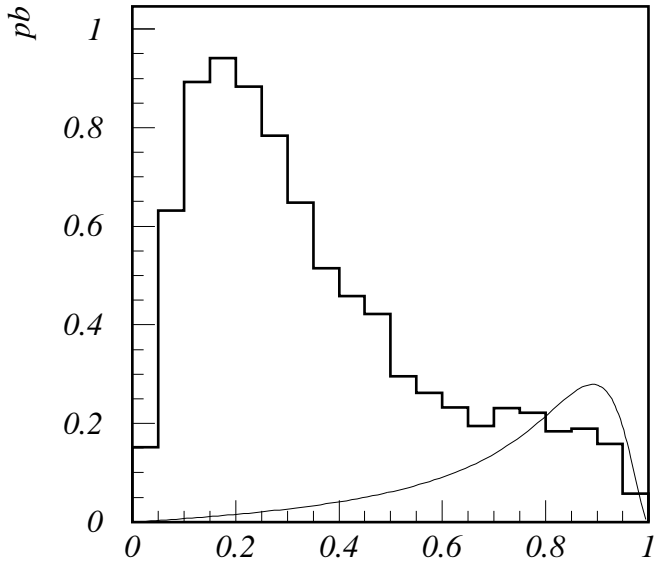
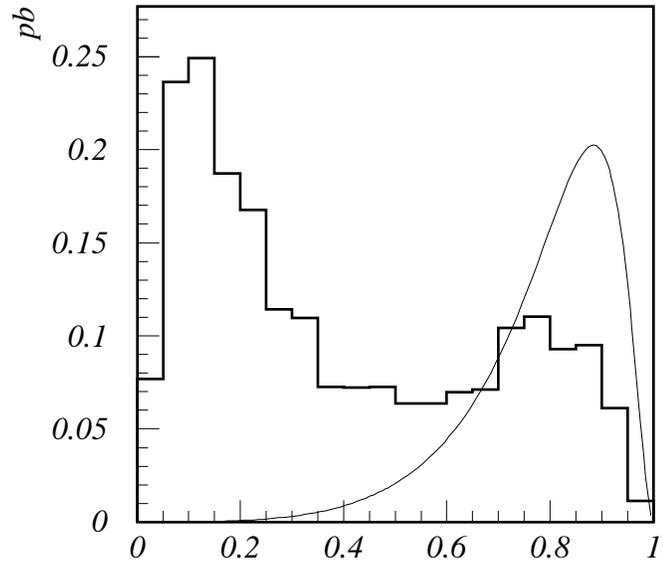
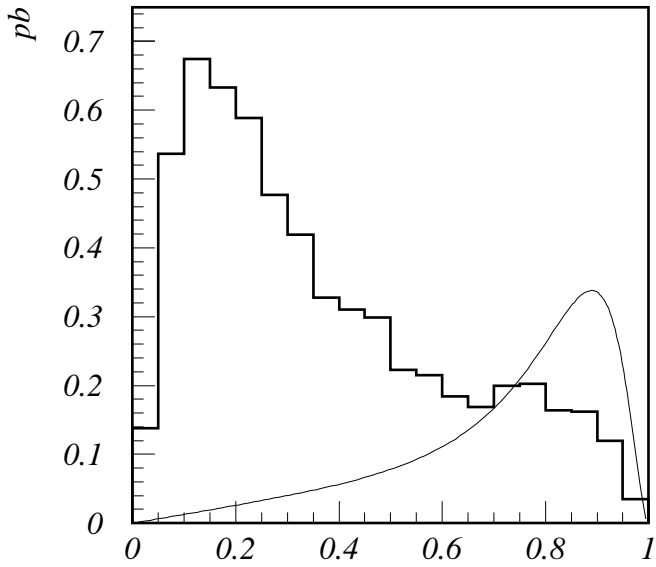
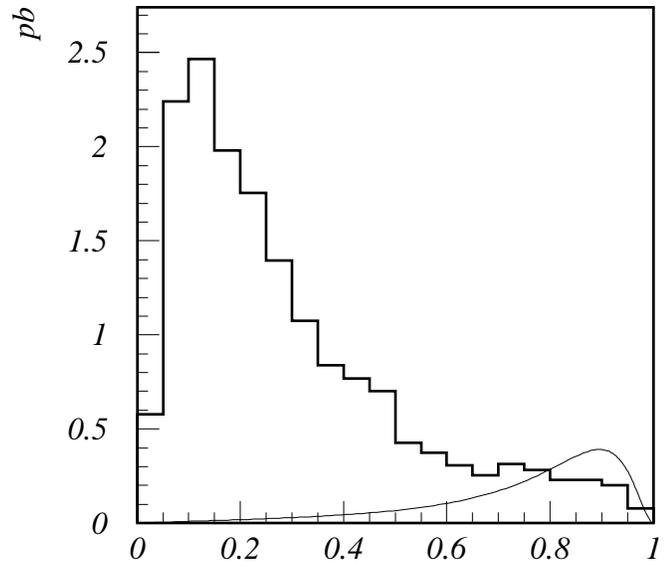
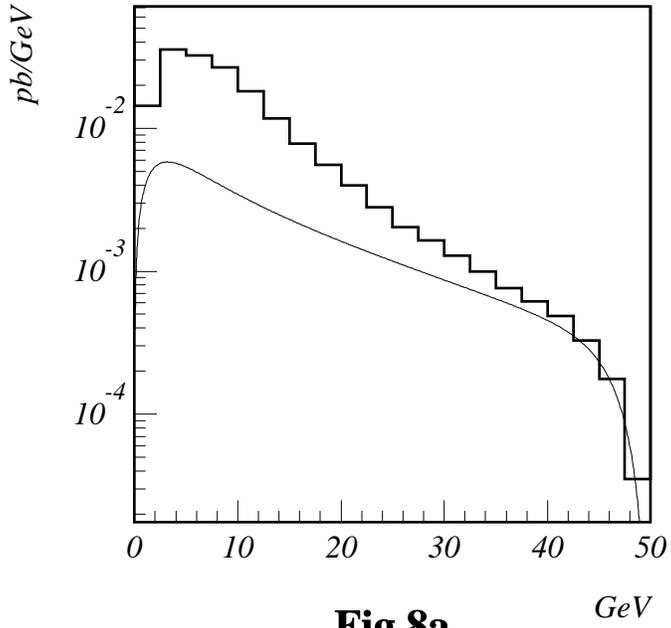
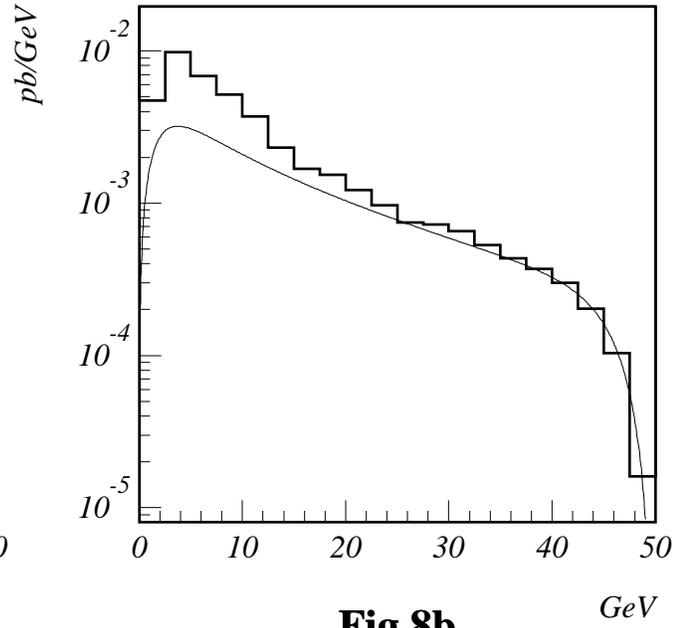
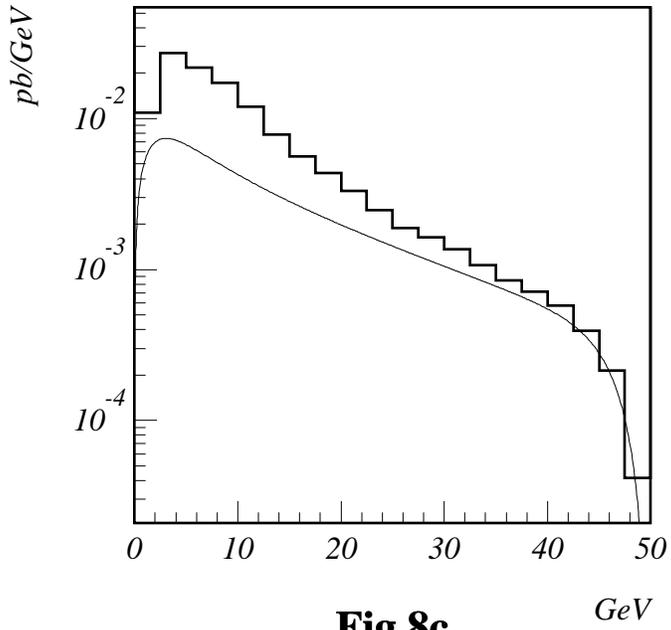
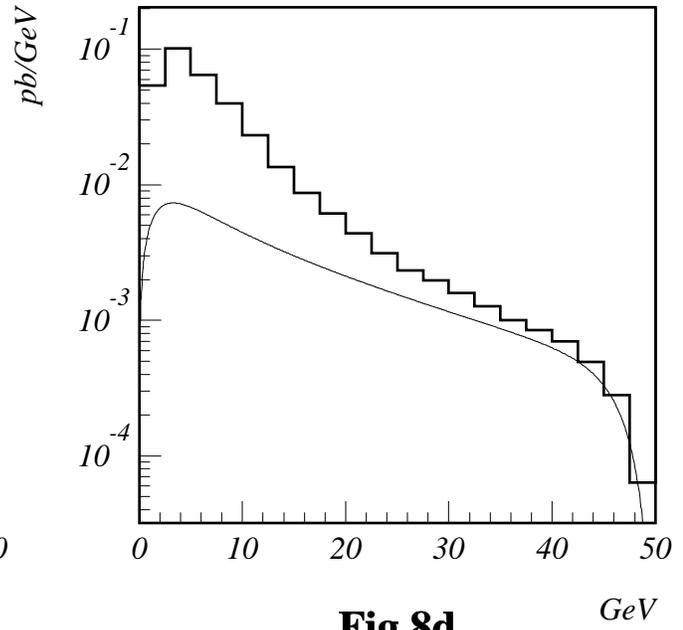
\end{document}